\newtheorem{theorem}{Theorem}

\newtheorem{proposition}[theorem]{Proposition}
\newtheorem{definition}[theorem]{Definition}
\newtheorem{example}[theorem]{Example}

\newtheorem{remark}[theorem]{Remark}

\documentclass[11pt]{article} 
\usepackage{amssymb}
\usepackage{sw20aip}  

\def\proof{\noindent{\bfseries Proof. }}

\def\endproof{\mbox{\ \rule{.1in}{.1in}}}
\def\limfunc#1{\mathop{\rm #1}}
\long
\def\TeXButton#1#2{#2}
\def\1{{\rm{\bf 1}}}

\def\C{{\rm{\bf C}}}
\def\Ci{{\rm{\bf C}}}

\def\N{{\rm{\bf N}}}
\def\Ni{{\rm{\bf N}}}

\def\R{{\rm{\bf R}}}
\def\Ri{{\rm{\bf R}}}

\input tcilatex

\begin{document}

\title{Feynman integrals for a class of exponentially growing potentials}
\author{\textbf{Tobias Kuna} \\
Universit\"at Bielefeld, D 33615 Bielefeld, Germany \and \textbf{Ludwig
Streit} \\
Universit\"at Bielefeld, D 33615 Bielefeld, Germany\\
CCM, Universidade da Madeira, P 9000 Funchal, Portugal \and \textbf{Werner
Westerkamp} \\
Universit\"at Bielefeld, D 33615 Bielefeld, Germany}
\date{}
\maketitle

\noindent
{\bf PAC:} 03.65.-w, 02.30.Mv, 02.50.Fz.

\medskip
\noindent
Published in: J. Math. Physics 39 (1998) p.4476-4491
\medskip

\begin{abstract}
We construct the Feynman integrands for a class of exponentially growing
time-dependent potentials as white noise functionals. We show that they
solve the Schr\"odinger equation. The Morse potential is considered as a
special case.
\end{abstract}

\tableofcontents

\thispagestyle{empty}

\newpage

\QSubDoc{Include intro}{
\LaTeXparent{C:/TKUNA/TEX;master}
                      
\ChildDefaults{chapter:0,page:1}

\section{INTRODUCTION}

As an alternative approach to quantum mechanics Feynman introduced the
concept of path integrals,\cite{FeHi65} which developed into an
extremely useful tool in many branches of theoretical physics.

Unfortunately Feynman's intuitive idea of averaging over some set of paths
is mathematically meaningful only for the heat equation where the underlying
structure for the free motion is based on the Brownian paths with Wiener
measure. This is stated by the famous Feynman Kac formula. To write down
solutions of the Schr\"odinger equation as path integrals is much more
involved and often less direct. A measure does not exist for the Feynman
integral as in the Euclidean case.\cite{Ca60} We do not give a full
list of references, but we would like to refer to Ref.~3,
\nocite{AlHK76} their method using infinite dimensional Fresnel integrals and their extensive list
of references to further approaches to Feynman integrals. Further we want to
mention one of the most widely used methods, analytic continuation. The
calculation itself takes place on the Euclidean side, where we have the
whole machinery of probabilistic theory at our disposal. The final step is
to perform an analytic continuation in some parameter such as time or mass
to get solutions of the Schr\"odinger equation. In Subsection
\ref{SeUnbounded} \ref{SuSeMass} 
we will prove that continuation in the mass is incompatible with
perturbation theory in the case we are considering. Additionally we want to
mention the method of Doss,\cite{Do80} using the Feynman Kac formula
and complex scaling. The method of Doss can also be used in the framework of
white noise analysis.\cite{HKPS93,W95}

White noise analysis,\cite{HKPS93,Ob94,Kuo96} 
is a framework which offers various generalizations of concepts known from
finite dimensional analysis to the infinite dimensional case, among them are
differential operators and Fourier transform.

The underlying random variable is not Brownian motion but rather its
velocity, white noise. Being independent at each time, white noise provides
a suitable infinite dimensional coordinate system. The ''integral'' is
understood as the dual pairing of a distribution with a test function, so
that the Feynman integrand itself has meaning as a distribution. This allows
us to calculate not only the propagator but, more generally, time ordered
expectation values. Important for the usefulness of any approach to Feynman
integrals is the class of potentials we are able to handle. In white noise
analysis Feynman integrals have been constructed for different classes of
potentials. The first were proposed by Ref.~9 \nocite{FPS91} 
and by Khandekar and Streit.\cite{KS92} This latter construction 
was generalized in Refs.~11 and 12  \nocite{LLSW93} \nocite{GKSS96}
to a wider class, allowing also time-dependent
interactions. Potentials there were given as superpositions of $\delta $%
-functions. Unfortunately this is restricted to one space dimension. 
In Ref.~6 \nocite{W95} 
another set was considered, the so called Albeverio-H\o egh-Krohn
class \cite{AlHK76} of potentials that are Fourier transforms of measures.
Here the space dimension is arbitrary, on the other hand the potentials are
smooth and bounded. We shall instead consider Laplace transforms of
measures, again for arbitrary finite space dimension. In Section~\ref
{SeUnbounded} we construct the Feynman integrand as a white noise
distribution and show that the corresponding propagator solves the
Schr\"odinger equation. The potentials are smooth but they grow in general
exponentially at $\pm \infty $. They are too singular to be handled by
Kato-Rellich perturbation theory. Nevertheless we show that the propagator
is analytic in the coupling constant and we write it as a Dyson series. In
Section \ref{SeMorse} we consider the special case of Morse potentials $%
V\left( x\right) =g(e^{2ax}-be^{ax})$ for illustration and for more explicit
calculations. This problem is solvable in closed form.
\cite{Kl90,PaSo84,CaInWi83,FiLeMu92} There the authors derive the
Green function, the spectrum and the eigenfunctions. These quantities are in
general not analytic in $g$. If we change from positive to negative $g$ we
also lose the essential self-adjointness of the corresponding Hamilton
operator. This dramatic change however does not destroy the analyticity of
the propagator.}

\QSubDoc{Include wn}{
\LaTeXparent{C:/TKUNA/TEX;master}
                      
\ChildDefaults{chapter:0,page:1}

\section{WHITE NOISE ANALYSIS}

In this section we give a brief overview of concepts and theorems of white
noise analysis which we use.\cite{HKPS93,Kuo96,Ob94}.

The starting point of $d$-dimensional white noise analysis is the real
separable Hilbert space 
\begin{equation}
L_d^2:=L^2\left( {\R}\right) \otimes {\R}^d,\quad d\in {\N,}
\end{equation}
which is unitary isomorphic to a direct sum of $d$ identical copies of $L^2({%
\R)}$ the space of real valued square-integrable functions with respect to
Lebesgue measure. The norm in $L_d^2$ is given by 
\begin{equation}
\left| f\right| _0:=\sum_{j=1}^d\int_{{\Ri}}f_j^2\left( s\right) \mathrm{d}s,%
\text{\quad }f\in L_d^2
\end{equation}
In this space we choose the densely imbedded nuclear space 
\begin{equation}
S_d:=S\left( {\R}\right) \otimes {\R}^d{.}
\end{equation}
A typical element $\xi \in S_d$ is a $d$-dimensional vector where each
component is a Schwartz test function. By $\left| \cdot \right| _p$ we
denote a family of Hilbert norms topologizing $S_d$. Together with the dual
space 
\begin{equation}
S_d^{\prime }:=S^{\prime }\left( {\R}\right) \otimes {\R}^d
\end{equation}
we obtain the basic nuclear triple 
\begin{equation}
S_d\subset L_d^2\subset S_d^{\prime }.
\end{equation}
Let $\mathcal{B}$ be the $\sigma $-algebra generated by cylinder sets on $%
S_d^{\prime }$. Using Minlos' theorem we construct a measure space on $%
\left( S_d^{\prime },\mathcal{B}\right) $ by fixing the characteristic
function.

\begin{equation}
\int_{S_d^{\prime }}\exp \left( i\left\langle \omega ,\xi \right\rangle
\right) \mathrm{d\mu }\left( \omega \right) =\exp \left( -\frac 12\left| \xi
\right| _0^2\right) ,\quad \xi \in S_d,
\end{equation}
where $\left\langle \cdot ,\cdot \right\rangle $ denotes the dual pairing
between $S_d^{\prime }$ and $S_d$. The space $\left( S_d^{\prime },\mathcal{B%
},\mu \right) $ is called the vector valued white noise space. Within this
formalism a version of $d$-dimensional Wiener's Brownian motion is given by 
\begin{equation}
B\left( \tau \right) :=\left\langle \omega ,{\1}_{\left[ 0,\tau \right)
}\right\rangle :=\left( \left\langle \omega ,{\1}_{\left[ 0,\tau \right)
}\otimes e_1\right\rangle ,\ldots ,\left\langle \omega ,{\1}_{\left[ 0,\tau
\right) }\otimes e_d\right\rangle \right) ,\quad \omega \in S_d^{\prime },
\label{DeBrownian}
\end{equation}
where $\{e_1,\ldots ,e_d\}$ denotes the canonical basis of ${\R}^d$. We
shall construct a Gel'fand triple with smooth and generalized functions of
white noise around the complex Hilbert space 
\begin{equation}
L^2\left( \mu \right) :=L^2\left( S_d^{\prime },\mathcal{B},\mu \right)
\end{equation}
and we denote the scalar product in this space by 
\begin{equation}
\left( \!\left( f,g\right) \!\right) =\int_{S_d^{\prime }}\bar f\left(
\omega \right) g\left( \omega \right) \mathrm{d}\mu \left( \omega \right)
,\quad \;f,g\in L^2\left( \mu \right) .
\end{equation}
We proceed by choosing first a special subspace $\left( S_d\right) ^1$ of
test functionals. Then we construct the Gel'fand triple 
\begin{equation}
\left( S_d\right) ^1\subset L^2\left( \mu \right) \subset \left( S_d\right)
^{-1}.
\end{equation}
Elements of the space $\left( S_d\right) ^{-1}$ are called Kondratiev
distributions, the well known Hida distributions form a subspace. Instead of
reproducing the explicit construction here,\cite{Ko78,KLS96}
we shall characterize the distributions by their $T$-transforms in
Theorem \ref{ThCharKonrad} below. Let $\Phi \in \left( S_d\right) ^{-1}$
then there exist $p,q\in {\N}_0$ such that we can define for every 
\begin{equation}
\xi \in U_{p,q}:=\left\{ \xi \in S_d\,\left| \,\left| \xi \right|
_p^2<2^{-q}\right. \right\}
\end{equation}
the $T$-transform by 
\begin{equation}
T\Phi \left( \xi \right) :=\left\langle \!\left\langle \Phi ,\exp \left(
i\left\langle \cdot ,\xi \right\rangle \right) \right\rangle \!\right\rangle
\end{equation}
where $\left\langle \!\left\langle \cdot ,\cdot \right\rangle
\!\right\rangle $ denotes the bilinear extension of the scalar product of $%
L^2\left( \mu \right) $. The definition of the $T$-transform can be extended
via analytic continuation to the complexification of $S_d$ which we denote
by $S_{d,{\Ci}}$. Further we need the definition of holomorphy in a nuclear
space.\cite{Ba85}

\begin{definition}
\label{DeHolomorph}A function $F:U\rightarrow {\C}$ is holomorphic on an
open set $U\subseteq S_{d,{\Ci}}$ iff for all $\theta _0\in U$

\begin{enumerate}
\item  for any $\theta \in S_{d,{\Ci}}$ the mapping $\lambda \mapsto F\left(
\theta _0+\lambda \theta \right) $ is holomorphic in some neighborhood of $0$
in ${\C}$,

\item  there exists an open neighborhood $U^{\prime }$ of $\theta _0$ such
that $F$ is bounded on $U^{\prime }$.
\end{enumerate}

$F$ is holomorphic at $0$ iff $F$ is holomorphic in a neighborhood of $0$.
\end{definition}

Now we can give the above mentioned characterization theorem, which is in
the case of Hida distributions due to Refs.~20 and 21,\nocite{PS91,KLPSW96}
and for Kondratiev distributions to Ref.~18.\nocite{KLS96}

\begin{theorem}
\label{ThCharKonrad}Let $U\subseteq S_{d,{\Ci}}$ be open and $F:U\rightarrow 
{\C}$ be holomorphic at zero, then there exists a unique $\Phi \in (S_d)^{-1}
$ such that $T\Phi =F$. Conversely, let $\Phi \in (S_d)^{-1}$ then $T\Phi $
is holomorphic at zero. The correspondence between $F$ and $\Phi $ is a
bijection if we identify holomorphic functions which coincide on an open
neighborhood of zero.
\end{theorem}

As a consequence of the characterization we have also a criterion for
sequences and integrals with respect to an additional parameter.%

\begin{theorem}
\label{ThSequenceKonrad}Let $\left( \Phi _n\right) _{n\in {\Ni}}$ be a
sequence in $(S_d)^{-1}$, such that there exists $U_{p,q}$, $p,q\in {\N}_0$,
so that

\begin{enumerate}
\item  all $T\Phi _n$ are holomorphic on $U_{p,q};$

\item  there exists $C>0$ such that $\left| T\Phi _n\left( \theta \right)
\right| \leq C$ for all $\theta \in U_{p,q}$ and all $n\in {\N}$,

\item  $\left( T\Phi _n\left( \theta \right) \right) _{n\in {\Ni}}$ is a
Cauchy sequence in ${\C}$ for all $\theta \in U_{p,q}$.
\end{enumerate}

\noindent Then $\left( \Phi _n\right) _{n\in {\Ni}}$ converges strongly in $%
(S_d)^{-1}$.
\end{theorem}

\begin{theorem}
\label{ThIntegralKonrad}Let $\left( \Lambda ,\mathcal{A},\nu \right) $ be a
measure space and $\lambda \mapsto \Phi _\lambda $ a mapping from $\Lambda $
to $(S_d)^{-1}$. We assume that there exists $U_{p,q}$ , $p,q\in {\N}_0$,
such that

\begin{enumerate}
\item  $T\Phi _\lambda $, is holomorphic on $U_{p,q}$ for every $\lambda \in
\Lambda $;

\item  the mapping $\lambda \mapsto T\Phi _\lambda \left( \theta \right) $
is measurable for every $\theta \in U_{p,q}$.

\item  there exists $C\in L^1\left( \Lambda ,\,\nu \right) $ such that 
\begin{equation}
\left| T\Phi _\lambda \left( \theta \right) \right| \leq C\left( \lambda
\right) 
\end{equation}
for all $\theta \in U_{p,q}$ and for $\nu $-almost all $\lambda \in \Lambda $%
.
\end{enumerate}

Then there exist $p^{\prime }$,$~q^{\prime }\in {\N}_0$, which only depend
on $p,q$, such that $\Phi _\lambda $ is Bochner integrable.

In particular, 
\begin{equation}
\int_\Lambda \Phi _\lambda \mathrm{d}\nu \left( \lambda \right) \in \left(
S_d\right) ^{-1}
\end{equation}

and $T\left[ \int_\Lambda \Phi _\lambda \mathrm{d}\nu \left( \lambda \right)
\right] $ is holomorphic on $U_{p^{\prime },q^{\prime }}$. We may
interchange dual pairing and integration 
\begin{equation}
\Bigg{\langle} \!\! \Bigg{\langle} \int_\Lambda \Phi_\lambda 
\mathrm{d}\nu \left( \lambda \right) ,\varphi \Bigg{\rangle}
\!\! \Bigg{\rangle} =\int_\Lambda \big{\langle} \! \big{\langle} \Phi _\lambda
,\varphi \big{\rangle} \! \big{\rangle} \mathrm{d}\nu \left( \lambda \right)
,\quad \varphi \in \left( S_d\right)^1.
\end{equation}
\end{theorem}

At the end of this section we want to give some examples of distributions.
\cite{HKPS93}

\begin{example}
\emph{To define the kinetic energy factor in the path integrals one would
like to give a meaning to the formal expression } 
\[
\exp \left( c\int_{{\Ri}}\omega \left( t\right) ^2\mathrm{d}t\right) ,
\]
\emph{where }$c$\emph{\ is a complex constant. We define the normalized
exponential } 
\[
\limfunc{Nexp}\left( c\int_{{\Ri}}\omega \left( t\right) ^2\mathrm{d}%
t\right) 
\]
\emph{as a distribution via the following }$T$\emph{-transform } 
\[
T\limfunc{Nexp}\left( c\int_{{\Ri}}\omega \left( t\right) ^2\mathrm{d}%
t\right) \left( \xi \right) =\exp \left( \frac 1{4c-2}\int_{{\Ri}}\xi
^2\left( \tau \right) \mathrm{d}\tau \right) ,\quad c\neq \frac 12.
\]
\end{example}

\begin{example}
\emph{Donsker's delta function\label{DeDonsker}. In order to `pin' Brownian
motion at a point }$a\in {\R}^d$\emph{\ we want to consider the formal
composition of the Dirac delta distribution with Brownian motion: }$\delta
\left( B\left( t\right) -a\right) .$\emph{\ This can be given a precise
meaning as a Hida distribution.\cite{Kuo92} Its }$T$
\emph{-transform is given by} 
\[
T\left[ \delta \left( B\left( t\right) -a\right) \right] \left( \xi \right)
=\frac 1{\left( 2\pi t\right) ^{\frac d2}}\exp \left( -\frac 1{2t}\left(
\int_0^ti\xi \left( s\right) ds+x_0-a\right) ^2-\frac 12\left| \xi \right|
_0^2\right) 
\]
\end{example}}

\QSubDoc{Include free}{
`\LaTeXparent{C:/TKUNA/TEX;master}
                      
\ChildDefaults{chapter:0,page:1}

\section{THE FREE FEYNMAN INTEGRAND\label{SeFree}}

\noindent We follow Refs.~9 and 23 \nocite{FPS91,HS83} in viewing the Feynman
integral as a weighted average over Brownian paths. We use a slight change
in the definition of the paths, which are here modeled by 
\begin{equation}
x\left( \tau \right) =x-\sqrt{\frac \hbar m}\int_\tau ^t\omega \left( \sigma
\right) \mathrm{d}\sigma :=x-\sqrt{\frac \hbar m}\!\left\langle \omega ,{\1}%
_{\left( \tau ,t\right] }\right\rangle   \label{DePath}
\end{equation}
In the sequel we set $\hbar =m=1$ unless otherwise stated. Correspondingly
we define the Feynman integrand for the free motion by \cite{LLSW93} 
\begin{equation}
I_0\left( x,t\mid x_0,t_0\right) =\limfunc{Nexp}\left( \frac{i+1}2\int_{{\Ri}%
}\omega ^2\left( \tau \right) \mathrm{d}\tau \right) \delta \left( x\left(
t_0\right) -x_0\right). 
\end{equation}
We recall that the delta distribution $\delta \left(
x\left( t_0\right) -x_0\right) $ is used to fix the starting point and plays
the role of an initial distribution. In the sequel instead of $I_0\left(
x,t\mid x_0,t_0\right) $ we will often use the shorthand $I_0$. Thus we get
for the $T$-transform

\begin{eqnarray}
TI_0\left( \xi \right) &=&\frac 1{\left( 2\pi i\left| t-t_0\right| \right)
^{\frac d2}}\exp \left[ -\frac i2\int_{{\R}}\xi ^2\left( \tau \right) 
\mathrm{d}\tau \right.  \label{TFeynmanExplicit} \\[3ex]
&&\left. -\frac 1{2i\left| t-t_0\right| }\left( \int_{t_0}^t\xi \left( \tau
\right) \,\mathrm{d}\tau +x-x_0\right) ^2\right] ,  \nonumber
\end{eqnarray}
Not only the expectation but also the $T$-transform has a physical meaning.
Integrating formally by parts we find 
\begin{eqnarray}
TI_0\left( \xi \right) &=&\int_{\mathcal{S}_d^{\prime }}I_0\left( \omega
\right) \exp \left( -i\int_{t_0}^tx\left( \tau \right) \cdot \stackrel{%
_{\bullet }}{\xi }\left( \tau \right) \mathrm{d}\tau \right) \mathrm{d}\mu
\left( \omega \right) \\[3ex]
&&\times \exp \left( -\frac i2\int_{\left[ t_0,t\right] ^c}\xi ^2\left( \tau
\right) \mathrm{d}\tau \right) \exp \left[ ix\cdot \xi \left( t\right)
-ix_0\cdot \xi \left( t_0\right) \right] .  \nonumber
\end{eqnarray}
The multiplication denoted by dot is just the scalar product in ${\R}^d$.
Indeed it is straightforward to verify that 
\begin{eqnarray}
K_0^{(\xi )}\left( x,t|x_0,t_0\right) &=&TI_0\left( \xi \right)
\label{FeynmanCorrection} \\[3ex]
&&\times \exp \left( \frac i2\int_{\left[ t_0,t\right] ^c}\xi ^2\left( \tau
\right) \mathrm{d}\tau \right) \exp \left[ ix_0\cdot \xi \left( t_0\right)
-ix\cdot \xi \left( t\right) \right]  \nonumber
\end{eqnarray}
\noindent   is the Green function corresponding to the potential $W=\,\, 
\stackrel{_{\bullet }}{\xi }\!\left( t\right) \cdot x$, i.e., it obeys the
Schr\"odinger equation 
\begin{equation}
\left( i\partial _t+\frac 12\triangle _d-\stackrel{_{\bullet }}{\xi }%
\!\left( t\right) \cdot x\right) K_0^{(\xi )}\left( x,t|x_0,t_0\right) =0
\label{schrodinger}
\end{equation}
with the initial condition 
\[
\lim_{t\searrow t_0}K_0^{(\xi )}\left( x,t|x_0,t_0\right) =\delta \left(
x-x_0\right) . 
\]}

\QSubDoc{Include class}{
\LaTeXparent{C:/TKUNA/TEX;master}
\ChildStyles{amssymb} 
\ChildDefaults{chapter:0,page:1}

\section{THE FEYNMAN INTEGRAND FOR A NEW CLASS OF UNBOUNDED POTENTIALS\label
{SeUnbounded}}

Now we construct the Feynman integrand for a new class of potentials and
calculate the propagators. In Subsection \ref{SuSeSG} we show that the
propagators solve the corresponding Schr\"odinger equation and in Subsection \ref
{SuSeTimeDep} we generalize to time-dependent potentials.

\subsection{The interactions}

\begin{definition}
\label{DePotential}Let $m$ be a complex measure on the Borel sets on ${\R}^d,
$ $d\geq 1$ fulfilling the following condition 
\begin{equation}
\int_{{\Ri}^d}e^{C\left| \alpha \right| }\,\mathrm{d\!}\left| m\right|
\!\left( \alpha \right) <\infty ,\quad \forall C>0.  \label{condition}
\end{equation}
We define a potential $V$ on ${\R}^d$ by 
\begin{equation}
V\left( x\right) =\int_{{\Ri}^d}e^{\alpha \cdot x}\mathrm{d}m\!\left( \alpha
\right) .
\end{equation}
\end{definition}

\begin{remark}
\emph{A consequence of the above condition (\ref{condition}}) \emph{is that
the measure }$m$\emph{\ is finite. By Lebesgue's dominated convergence
theorem we obtain that the potentials are restrictions to the real line of
entire functions. In particular they are locally bounded and without
singularities. However they are in general unbounded at }$\pm \infty $\emph{.%
}
\end{remark}

\begin{remark}
\emph{The time-dependent case will be considered in Subsection \ref
{SuSeTimeDep}.}
\end{remark}

\begin{example}
\emph{Every finite measure with compact support fulfills the above condition 
(\ref{condition}).}
\end{example}

\begin{example}
\emph{The simplest example is the Dirac measure in one dimension }$m\left(
\alpha \right) :=g\,\,\delta _a\left( \alpha \right) $\emph{\ for }$a>0$%
\emph{\ and }$g\in {\R}$\emph{. The associated potential is }$V\left(
x\right) =g\,e^{ax}$\emph{. Obviously all polynomials of exponential
functions of the above kind are also in our class, too, e.g. }$\sinh \left(
ax\right) $, $\cosh (ax)$\emph{.}
\end{example}

\begin{example}
\emph{In particular the well known Morse potential }$V(x):=g(
e^{-2ax}-2\gamma e^{-ax}) $ \emph{with} $g,a,x\in {\R}$\emph{\ and }$%
\gamma >0$\emph{\ is included in our class. We will discuss this potential
in Section \ref{SeMorse}} \emph{in more detail.}
\end{example}

\begin{example}
\emph{If we choose a Gaussian density, we get potentials of the form }$%
V\left( x\right) =ge^{bx^2}$\emph{\ with }$b,x\in {\R}$\emph{.}
\end{example}

\begin{example}
\emph{Further entire functions of arbitrary high order of growth are inside
of our class. More explicitly, the measures }$m\left( \alpha \right) :=\Theta
\left( \alpha \right) \exp \left( -k\alpha ^{1+b}\right) $\emph{\ with }$b$%
\emph{, }$k>0$\emph{\ and }$x\in {\R}$\emph{\ fulfill the condition (\ref
{condition}). The corresponding potentials are entire functions of order }$%
1+ 1/b$\emph{, see Ref.~24 \nocite{Lu70} Lemma 7.2.1.}
\end{example}

\subsection{The Feynman integrand as a generalized white noise functional}

In order to handle potentials of the form given above 
\[
V\left( x\right) =\int_{{\Ri}^d}e^{\alpha \cdot x}\mathrm{d}m\!\left( \alpha
\right) 
\]
within our approach we must give a meaning to the following pointwise
multiplication 
\begin{equation}
I=I_0\cdot \exp \left( -i\int_{t_0}^tV\left( x\left( \tau \right) \right) 
\mathrm{d}\tau \right) \text{,}
\end{equation}
where 
\[
x\left( \tau \right) =x-\int_\tau ^t\omega \left( s\right) \,\mathrm{d}s 
\]
is a path, as in Section \ref{SeFree}. For convenience we assume $t_0<t$
in the sequel. As a first step we formally expand the exponential 
\begin{equation}
I=\sum_{n=0}^\infty \frac{\left( -i\right) ^n}{n!}\int_{[t_0,t]^n}\int_{{\Ri}%
^{dn}}I_0\cdot \prod_{j=1}^ne^{\alpha _j\cdot x\left( \tau _j\right)
}\prod_{j=1}^n\mathrm{d}m\!\left( \alpha _j\right) \mathrm{d}^n\tau
\end{equation}
into a perturbation series. In Theorem \ref{ThSG} we will show the existence
of the integrals and the series . But first we have to give a definition for
the product 
\begin{equation}
I_0\cdot \prod_{j=1}^ne^{\alpha _j\cdot x\left( \tau _j\right) }
\end{equation}
Products of this type have already been considered, see\textbf{\ }e.g.
Example \ref{DeDonsker}. In view of the Characterization Theorem \ref
{ThCharKonrad} it is enough to define the product via its $T$-transform.
Arguing formally we obtain 
\begin{eqnarray}
\lefteqn{T\left( I_0\cdot \prod_{j=1}^ne^{\alpha _j\cdot x\left( \tau _j\right)
}\right) \!\!\left( \xi \right)} \\[3ex]
&=&\int_{\mathcal{S}_d^{\prime }}I_0\cdot \prod_{j=1}^ne^{\alpha _j\cdot
x\left( \tau _j\right) }\exp \left( i\left\langle \omega ,\xi \right\rangle
\right) \mathrm{d}\mu \!\left( \omega \right)  \nonumber \\[3ex]
&=&TI_0\left( \xi +i\sum_{j=1}^n\alpha _j{\1}_{\left[ \tau _j,t\right)
}\right) \exp \left( \sum_{j=1}^n\alpha _j\cdot x\right) .  \nonumber
\end{eqnarray}
We only need to verify that $TI_0$ is extendable to $\xi
+i\sum_{j=1}^n\alpha _j{\1}_{\left( \tau _j,t\right] }.$ This is clearly
fulfilled, since the explicit formula (\ref{TFeynmanExplicit}) extends
continuously to all $\xi \in L_d^2.$

Hence we may define the product in this way:

\begin{proposition}
\label{PrProduct}Let $\tau _j\in [t_0,t]$ for $j=1,\ldots ,n$; $t_0<t$ and $%
\alpha _j\in {\R}^d$. Then the pointwise product 
\begin{equation}
\Phi _n=I_0\cdot \prod_{j=1}^ne^{\alpha _j\cdot x\left( \tau _j\right) }
\label{Product}
\end{equation}
defined by 
\begin{eqnarray*}
T\Phi _n\left( \xi \right) &=& TI_0\left( \xi +i\sum_{j=1}^n\alpha _j{\1}%
_{\left( \tau _j,t\right] }\right) \exp \left( \sum_{j=1}^n\alpha _j\cdot
x\right)  \\[3ex]
\ &=&\left( 2\pi i\left( t-t_0\right) \right) ^{-d/2}\exp \left[ -\frac
i2\int_{{\R}}\left( \xi \left( s\right) +i\sum_{j=1}^n\alpha _j{\1}_{\left(
\tau _j,t\right] }\left( s\right) \right) ^2\mathrm{d}s\right]  \\[3ex]
& &\times \exp \left\{ -\frac 1{2i\left( t-t_0\right) }\left[ \int_{t_0}^t\xi
\left( s\right) \mathrm{d}s+i\sum_{j=1}^n\alpha _j\left( t-\tau _j\right)
+\left( x-x_0\right) \right] ^2\right\}  \\[3ex]
& &\times \exp \left( \sum_{j=1}^n\alpha _j\cdot x\right) 
\end{eqnarray*}
is a Kondratiev distribution.
\end{proposition}

\TeXButton{Proof}{\proof}Obviously this has an extension in $\xi \in 
\mathcal{S}_d({\R)}$ to all $\theta \in \mathcal{S}_{d,{\Ci}}({\R)}$ and
fulfills the first part of Definition \ref{DeHolomorph}. In order to prove
that $\Phi _n\in \left( \mathcal{S}_d\right) ^{-1}$ by applying Theorem \ref
{ThCharKonrad}, we need a bound 
\begin{eqnarray}
\lefteqn{ \left| T\Phi _n\left( \theta \right) \right|}  \nonumber \\*[3ex]
\ &\leq &\left( 2\pi \left( t-t_0\right) \right) ^{-d/2}\exp \left\{ \frac
12\left| \theta \right| _0^2+\int_{\text{I\negthinspace R}}\left| \theta
\left( s\right) \sum_{j=1}^n\alpha _j{\1}_{\left( \tau _j,t\right] }\left(
s\right) \right| \mathrm{d}s\right\}  \nonumber \\*[3ex]
&&\times \exp \left\{ \frac 1{2\left( t-t_0\right) }\left[ 2\left|
\left( \left( x-x_0\right) +\sum_{j=1}^n\alpha _j\left( t-\tau _j\right)
\right) \int_{t_0}^t\theta \left( s\right) \mathrm{d}s\right| \right. \right.
\nonumber \\*[3ex]
&&\ \ \ \ \ \left. \left. +\left( t-t_0\right) \left| \theta \right|
_0^2+2\left| \left( x-x_0\right) \cdot \sum_{j=1}^n\alpha _j\left( t-\tau
_j\right) \right| \right] \right\} \exp \left( \sum_{j=1}^n\alpha _j\cdot
x\right)  \nonumber \\[3ex]
\ &\leq &\left( 2\pi \left( t-t_0\right) \right) ^{-d/2}\exp \left(
\sum_{j=1}^n\left| \alpha _j\right| \left| x-x_0\right| +\sum_{j=1}^n\left|
\alpha _j\right| \left| x_0\right| \right)  \label{estimate} \\*[3ex]
&&\times \exp \left[ \left| \theta \right| _0^2+\left( 2\sqrt{t-t_0}%
\sum_{j=1}^n\left| \alpha _j\right| +\frac{\left| x-x_0\right| }{\sqrt{t-t_0}%
}\right) \left| \theta \right| _0\right]  \nonumber \\*[3ex]
\ &=:& C_n\left( \alpha _1,\ldots ,\alpha _j,\theta \right) .  \nonumber
\end{eqnarray}
Thus\textbf{\ $\Phi _n$ }is a Kondratiev distribution, in fact by the above
bound it is also a Hida distribution.\cite{KLPSW96} \hfill 
\endproof

Now we are able to prove the existence of the integrand.

\begin{theorem}
\label{ThDistrI}Let $V$ be as in Definition \ref{DePotential}. Then 
\begin{equation}
I:=\sum_{n=0}^\infty \frac{\left( -i\right) ^n}{n!}\int_{\left[ t_0,t\right]
^n}\int_{{\Ri}^{dn}}I_0\cdot \prod_{j=1}^ne^{\alpha _j\cdot x\left( \tau
_j\right) }\prod_{j=1}^n\mathrm{d}m\!\left( \alpha _j\right) \mathrm{d}%
^n\tau 
\end{equation}
exists as a generalized white noise functional. The series converges in the
strong topology of $(\mathcal{S}_d)^{-1}$. The integrals exist in the sense
of Bochner integrals. Therefore we can express the $T$-transform by 
\begin{equation}
TI\left( \theta \right) =\sum_{n=0}^\infty \frac{\left( -i\right) ^n}{n!}%
\int_{\left[ t_0,t\right] ^n}\int_{{\Ri}^{dn}}T\left( I_0\cdot
\prod_{j=1}^ne^{\alpha _j\cdot x\left( \tau _j\right) }\right) \!\left(
\theta \right) \prod_{j=1}^n\mathrm{d}m\!\left( \alpha _j\right) \mathrm{d}%
^n\tau 
\end{equation}

for all $\theta $ in a neighborhood of zero 
\begin{equation}
U_{p,q}:=\left\{ \theta \in S_{d,}{}_{{\Ci}}\left| \,\,2^q\left| \theta
\right| _p<1\right. \right\}   \label{UPQ}
\end{equation}
for some $p,q\in {\N}_0 $.
\end{theorem}

\proof 
We have already shown in Proposition \ref{PrProduct} that the product 
\[
\Phi _n:=I_0\cdot \prod_{j=1}^ne^{\alpha _j\cdot x\left( \tau _j\right) } 
\]
is a Kondratiev distribution, moreover, we derived the estimate (\ref
{estimate}). In order to see that the integrals exist in the sense of
Bochner we want to apply Theorem \ref{ThIntegralKonrad}. As the $T$%
-transform of $\Phi _n$ is entire in $\xi \in \mathcal{S}_{d,{\Ci}}({\R)}$
and measurable, it remains only to derive a suitable bound 
\begin{eqnarray}
\lefteqn{ \int_{\left[ t_0,t\right] ^n}\int_{{\Ri}^{dn}}C_n(\alpha _1,\ldots
,\alpha _j,\theta )\prod_{j=1}^n\mathrm{d\!}\left| m\right| \!\left( \alpha
_j\right) \mathrm{d}^n\tau}  \label{EstimateIntegral} \\[3ex]
\ &\leq &\left( 2\pi \left( t-t_0\right) \right) ^{-d/2}(t-t_0)^n\exp \left(
\left| \theta \right| _0^2+\frac{\left| x-x_0\right| }{\sqrt{t-t_0}}\left|
\theta \right| _0\right)  \nonumber \\[3ex]
&&\times \,\left\{ \int_{{\Ri}^d}\exp \left( \left[ \left|
x-x_0\right| +\left| x_0\right| +2\sqrt{t-t_0}\left| \theta \right|
_0\right] \left| \alpha \right| \right) \,\,\mathrm{d\!}\left| m\right|
\!\,\left( \alpha \right) \right\} ^n  \nonumber
\end{eqnarray}
which is finite since the measure satisfies condition (\ref{condition}). Due
to Theorem \ref{ThIntegralKonrad} there exists an open neighborhood $%
U {\ }${independent} of $n$ and 
\[
I_n:=\int_{\left[ t_0,t\right] ^n}\int_{{\Ri}^{dn}}\Phi _n\prod_{j=1}^n%
\mathrm{d}m\!\left( \alpha _j\right) \mathrm{d}^n\tau \in \left( \mathcal{S}%
_d\right) ^{-1},\quad \forall n\in {\N} 
\]
with $TI_n$ is holomorphic on $U$. To finish the proof we must
show that the series converges in $(\mathcal{S}_d)^{-1}$ in the strong
sense. For that we apply Theorem \ref{ThSequenceKonrad}. We know that $TI_n$
is holomorphic on $ U $ and we can bound it by 
\begin{eqnarray}
\lefteqn{ \left| TI\left( \theta \right) \right|}  \label{EqBound} \\[3ex]
\ &\leq &\sum_{n=0}^\infty \frac 1{n!}\left| TI_n\left( \theta \right)
\right|  \nonumber \\[3ex]
\ &\leq &\left( 2\pi \left( t-t_0\right) \right) ^{-d/2}\exp \left( \left|
\theta \right| _0^2+\frac{\left| x-x_0\right| }{\sqrt{t-t_0}}\left| \theta
\right| _0\right)  \nonumber \\[3ex]
&&\times \exp \left\{ \left( t-t_0\right) \int_{{\Ri}^d}\exp
\left[ \left( \left| x\right| +2\left| x_0\right| +2\sqrt{t-t_0}\left|
\theta \right| _0\right) \left| \alpha \right| \right] \,\,\,\mathrm{d\!}%
\left| m\right| \!\left( \alpha \right) \right\}  \nonumber \\[3ex]
\ &<&\infty  \nonumber
\end{eqnarray}
for $\theta \in U $, so that we prove $I\in \left( \mathcal{S}%
_d\right) ^{-1}$.\hfill\TeXButton{End Proof}{\endproof}

\begin{remark}
\emph{The bound established in the proof above has a trivial, but rather
surprising consequence. For the forthcoming discussion it is convenient to
show the dependence on the coupling constant explicitly, so that we get } 
\begin{equation}
TI\left( \theta \right) =\sum_{n=0}^\infty \frac{\left( -ig\right) ^n}{n!}%
\int_{\left[ t_0,t\right] ^n}\int_{{\Ri}^{dn}}T\left( I_0\cdot
\prod_{j=1}^ne^{\alpha _j\cdot x\left( \tau _j\right) }\right) \! \left( \theta
\right) \prod_{j=1}^n\mathrm{d}m\!\left( \alpha _j\right) \mathrm{d}^n\tau 
\end{equation}
\emph{which is a perturbation series in the coupling constant. Recall the
bound we have already calculated; we obtain that } 
\begin{eqnarray}
\lefteqn{\left| TI\left( \theta \right) \right|}  \\*[3ex]
&\leq& \sum_{n=0}^\infty \frac{\left| g\right| ^n}{n!}\left| TI_n\left( \theta
\right) \right| \leq \left( 2\pi \left( t-t_0\right) \right) ^{-d/2}\exp
\left( \left| \theta \right| _0^2+\frac{\left| x-x_0\right| }{\sqrt{t-t_0}}%
\left| \theta \right| _0\right)   \nonumber \\*[3ex]
&&\times \exp \left\{ \left| g\right| \left( t-t_0\right) \int_{{\Ri}^d}\exp
\left( \left[ \left| x-x_0\right| +\left| x_0\right| +2\sqrt{t-t_0}\left|
\theta \right| _0\right] \left| \alpha \right| \right) \mathrm{d\!}\left|
m\right| \!\left( \alpha \right) \right\}   \nonumber
\end{eqnarray}
\emph{and hence }$TI\left( \theta \right) $\emph{\ is entire in the coupling
constant }$g$\emph{\ for all fixed }$x,x_0,t_0<t$\emph{\ and }$\theta \in
U_{p,q}$\emph{. This is surprising, since the corresponding Hamilton
operators, even if they are essentially self-adjoint for }$g>0$\emph{,
lose this property for }$g<0$\emph{\ in general.} \emph{Quantities such as
eigenvalues and eigenvectors will not be analytic in the coupling constant.
(On the other hand under a stronger condition than (\ref{condition})
Albeverio et al. \cite{AlBrHa96} have shown that the solution of the
Schr\"odinger equation }$\Psi _t\left( x\right) $ \emph{is analytic in the
coupling constant if the initial wave function }$\Psi _0$ \emph{as a
function of }$x$\emph{\ is from a certain class of analytic functions.)}
\end{remark}

\subsection{Schr\"odinger equation\label{SuSeSG}}

Our aim is to prove that the propagator constructed above indeed solves the
Schr\"odinger equation. We are able to show that for $t_0<t$ the propagator
does not only solve it in the sense of distributions, but also in the sense
of ordinary functions. Similar to the free case, see equation (\ref
{FeynmanCorrection}), we can also give the test function in the $T$%
-transform a physical meaning corresponding to a time dependent homogeneous
external small force in the sense of (\ref{UPQ}). To compensate the extra
factors appearing in the formal integration by parts, see again (\ref
{FeynmanCorrection}), we consider 
\begin{equation}
TI\left( \theta \right) \cdot \exp \left[ \frac i2\int_{\left[ t_0,t\right]
^c}\theta ^2\left( s\right) \,\mathrm{d}s+ix_0\cdot \theta \left( t_0\right)
-ix\cdot \theta \left( t\right) \right].
\end{equation}
This then produces the Schr\"odinger propagator as follows
\begin{theorem}
\label{ThSG}Let $V$ be as in Definition \ref{DePotential}. Then 
\begin{eqnarray}
\lefteqn{K^{\left( \theta \right) }\left( x,t\mid x_0,t_0\right)} 
\label{PropagatorExplicit} \\[3ex]
&=&\sum_{n=0}^\infty \frac{\left( -ig\right) ^n}{n!}\left( 2\pi i\left(
t-t_0\right) \right) ^{-d/2}  \nonumber \\[3ex]
& &\times \int_{\left[ t_0,t\right] ^n}\int_{{\Ri}^{dn}}\exp \left\{ -\frac
i2\int_{{t}_0}^t\left( \theta \left( s\right) +i\sum_{j=1}^n\alpha _j{\1}%
_{\left( \tau _j,t\right] }\left( s\right) \right) ^2\mathrm{d}s\right\}  
\nonumber \\[3ex]
& &\times \exp \left\{ -\frac 1{2i\left( t-t_0\right) }\left[ \int_{t_0}^t\theta
\left( s\right) \mathrm{d}s+i\sum_{j=1}^n\alpha _j\left( t-\tau _j\right)
+\left( x-x_0\right) \right] ^2\right\}   \nonumber \\[3ex]
& &\times \exp \left( \sum_{j=1}^n\alpha _j\cdot x\right) \exp \left( ix_0\cdot
\theta \left( t_0\right) -ix\cdot \theta \left( t\right) \right)
\prod_{j=1}^n\mathrm{d}m\!\left( \alpha _j\right) \mathrm{d}^n\tau  
\nonumber
\end{eqnarray}
solves the Schr\"odinger equation for all $x,x_0,t_0<t$ 
\begin{equation}
\left( i\frac \partial {\partial t}+\frac 12\triangle _d-gV\left( x\right) -x%
\cdot \stackrel{_{\bullet }}{ \theta }\left( t\right) \right) K^{\left(
\theta \right) }\left( x,t\mid x_0,t_0\right) =0.
\end{equation}
with initial condition 
\begin{equation}
\lim_{t\searrow t_0}K^{\left( \theta \right) }\!\left( x,t\mid
x_0,t_0\right) =\delta \!\left( x-x_0\right) 
\end{equation}
\end{theorem}

\begin{remark}
We may also write the propagator as a product of the free propagator
with a perturbation series 
\begin{eqnarray}
\lefteqn{K^{\left( \theta \right) }\left( x,t\mid x_0,t_0\right)} 
\label{EqPropProduct} \\[3ex]
&=&K_0^{\left( \theta \right) }\left( x,t\mid x_0,t_0\right) \cdot
\sum_{n=0}^\infty \frac{\left( -ig\right) ^n}{n!}\left( t-t_0\right)
^n\int_{\left[ 0,1\right] ^n}\int_{{\Ri}^{dn}}\prod_{j=1}^n\mathrm{d}%
m\!\left( \alpha _j\right) \mathrm{d}^n\sigma   \nonumber \\[3ex]
&&\times \exp \left\{ -\frac i2\left( t-t_0\right) \left[
\sum_{j=1}^n\sum_{k=1}^n\alpha _j\cdot \alpha _k\left( \sigma _j\sigma _k
-\sigma _j \wedge \sigma_k \right) \right] \right\}   \nonumber
\\[3ex]
& &\times \exp \left\{ \sum_{j=1}^n\alpha _j\cdot \left[ \sigma _jx+\left(
1-\sigma _j\right) x_0+\sigma _j\int_{\left( t-t_0\right) \sigma
_j+t_0}^t\theta \left( s\right) \mathrm{d}s\right. \right.   \nonumber \\[3ex]
& &\ \ \ \ \ \left. \left. -(1-\sigma _j)\int_{t_0}^{\left( t-t_0\right) \sigma
_j+t_0}\theta \left( s\right) \mathrm{d}s\right] \right\}.   \nonumber
\end{eqnarray}
\end{remark}

\begin{remark}
The bounds in the following proof also yield that $\,$for fixed $\theta \in
U_{p,q}$ the above series as a function of $x,t,x_0,t_0$ is $C^\infty $ and
that we are allowed to interchange integration and summation with
differentiation.
\end{remark}

\TeXButton{Proof}{\proof} It is easy to see that $T\Phi _n\left( \theta
\right) $ is $C^\infty $ in the variables $x,x_0,t_0,t$ if $t<t_0$. For
simplification we introduce the following abbreviation 
\begin{eqnarray}
\lefteqn{\ K^{\left( \theta \right) }\left( x,t\mid x_0,t_0\right)} \\[3ex]
\ &=& \sum_{n=0}^\infty \frac{\left( -ig\right) ^n}{n!}K_n^{\left( \theta \right)
}\left( x,t\mid x_0,t_0\right)  \nonumber
\end{eqnarray}
where 
\begin{eqnarray}
\lefteqn{K_n^{\left( \theta \right) }\left( x,t\mid x_0,t_0\right)} \\[3ex]
\ &=&\int_{\left[ t_0,t\right] ^n}\int_{{\Ri}^{dn}}T\Phi _n\left( \theta
\right) \exp \left( \frac i2\int_{\left[ t_0,t\right] ^c}\theta ^2\left(
s\right) \,\mathrm{d}s+ix_0\cdot \theta \left( t_0\right) -ix\cdot \theta
\left( t\right) \right)  \nonumber \\[3ex]
&&\ \ \ \ \ \times \prod_{j=1}^n\mathrm{d}m\!\left( \alpha _j\right) \mathrm{d%
}^n\tau .  \nonumber
\end{eqnarray}
By direct computation we obtain 
\begin{eqnarray}
\left( i\frac \partial {\partial t}+\frac 12\triangle _d\right) \left[
T\Phi _n\left( \theta \right) \exp \left( \frac i2\int_{\left[ t_0,t\right]
^c}\theta ^2\left( s\right) \,\mathrm{d}s+ix_0\cdot \theta \left( t_0\right)
-ix\cdot \theta \left( t\right) \right) \right]  \nonumber   \\[3ex]
=x \cdot \stackrel{_{\bullet }}{\theta }\left( t\right) \,\left[ T\Phi
_n\left( \theta \right) \exp \left( \frac i2\int_{\left[ t_0,t\right]
^c}\theta ^2\left( s\right) \,\mathrm{d}s+ix_0\cdot \theta \left( t_0\right)
-ix\cdot \theta \left( t\right) \right) \right]  \label{EqPhin}
\end{eqnarray}
Thus the above product alone solves already the free Schr\"odinger equation
with an extra small external force. The next step is to calculate the
derivatives of $K_n^{\left( \theta \right) }$. We interchange them with the
integrals. Since the domain of the $\tau $-integrals depends on $t$, we get
an extra term compared to (\ref{EqPhin}) if we differentiate with respect to 
$t$. Using the theorem of Fubini we recognize that this term corresponds to
the potential $V\left( x\right) $. 
\begin{eqnarray}
\lefteqn{\left( i\frac \partial {\partial t}+\frac 12\triangle _d\right)
K_n^{\left( \theta \right) }\left( x,t\mid x_0,t_0\right)} \\[3ex]
&=&x \cdot \stackrel{_{\bullet }}{ \theta }\left( t\right) K_n^{\left(
\theta \right) }\left( x,t\mid x_0,t_0\right) +\frac{in}{g}V\left( x\right)
K_{n-1}^{\left( \theta \right) }\left( x,t\mid x_0,t_0\right)  \nonumber
\end{eqnarray}
which is a typical recursion formula driven by the potential. By summing up
we see that $K^{\left( \theta \right) }\left( x,t\mid x_0,t_0\right) $
solves formally the Schr\"odinger equation. It remains to justify the
operations above. This can be done similarly to the proof of Theorem \ref
{ThDistrI} if we bound the derivatives of $T\Phi _n\left( \theta \right) $
as follows. The derivatives of $T\Phi _n\left( \theta \right) $ have the
form 
\begin{eqnarray}
&& \left\{ a_0\left( \tau _1,\ldots ,\tau _n,\theta \right)
+a_1\left( \tau _1,\ldots ,\tau _n,\theta \right) \cdot \left(
\sum_{j=1}^n\alpha _j\right) \right. \\[3ex]
&&\ \ \ \ \left. +a_2\left( \tau _1,\ldots ,\tau _n,\theta \right) 
\cdot \left( \sum_{j=1}^n\alpha _j\right) ^2 \right\} 
\cdot T\Phi _n\left( \theta \right)
  \nonumber
\end{eqnarray}
where $a_i$ are continuous in the $\tau _j$. As the $\tau _j$ varies only in
a compact domain we can bound the derivatives by 
\begin{equation}
\left\{ b_0\left( \theta \right) +b_1\left( \theta \right) +2b_2\left(
\theta \right) \right\} \cdot \exp \left( \sum_{j=1}^n\left| \alpha
_j\right| \right) \cdot C_n(\alpha _1,\ldots ,\alpha _j,\theta ),
\end{equation}
where $C_n$ is the bound in the proof of Proposition \ref{PrProduct} and 
\begin{equation}
b_i\left( \theta \right) :=\sup\limits_{\tau _j\in \left[ t_0,t\right]
}\left| a_i\left( \tau _1,\ldots ,\tau _n,\theta \right) \right| .
\end{equation}
The rest of the proof can be done as before.\hfill\TeXButton{End Proof}
{\endproof}

\subsection{Continuation to imaginary mass\label{SuSeMass}}

By continuation to imaginary mass we obtain a formal perturbation series for
the propagator of the heat equation. If the measure in Definition 
\ref{DePotential} is positive we show that this perturbation series diverges. 
In the following discussion we need the explicit dependence of the
propagator on $\hbar $ and $m$. For simplification we put $\theta =0$%
. $K^{\left( 0\right) }$ is abbreviated as $K$. By (\ref{EqPropProduct}) we
obtain 
\begin{eqnarray}
\lefteqn{K\left( x,t\mid x_0,t_0\right)} \\[3ex]
\ &=&K_0\left( x,t\mid x_0,t_0\right) \cdot \sum_{n=0}^\infty \frac
1{n!}\left( \frac{-ig\left( t-t_0\right) }\hbar \right) ^n\int_{\left[
0,1\right] ^n}\int_{{\Ri}^{dn}}\prod_{j=1}^n\mathrm{d}m\!\left( \alpha
_j\right) \mathrm{d}^n\sigma  \nonumber \\[3ex]
&&\times \exp \left\{ -\frac{i\hbar }{2m}\left( t-t_0\right) \left[
\sum_{j=1}^n\sum_{k=1}^n\alpha _j\cdot \alpha _k\left(  \sigma _j\sigma _k
-\sigma _j \wedge \sigma_k  \right) \right] \right\}  \nonumber \\[3ex]
&&\times \exp \left\{ \sum_{j=1}^n\alpha _j\cdot \left[ \sigma _jx+\left(
1-\sigma _j\right) x_0\right] \right\}.  \nonumber
\end{eqnarray}
If we perform a formal analytic continuation in the mass from $m$ to $im$ we
get 
\begin{eqnarray}
\lefteqn{K^H\left( x,t\mid x_0,t_0\right)} \\[3ex]
\ &=&K_0^H\left( x,t\mid x_0,t_0\right) \cdot \sum_{n=0}^\infty \frac
1{n!}\left( \frac{-ig\left( t-t_0\right) }\hbar \right) ^n\int_{\left[
0,1\right] ^n}\int_{{\Ri}^{dn}}\prod_{j=1}^n\mathrm{d}m\!\left( \alpha
_j\right) \mathrm{d}^n\sigma  \nonumber \\[3ex]
&& \times \exp \left\{ -\frac \hbar {2m}\left( t-t_0\right) \left[
\sum_{j=1}^n\sum_{k=1}^n\alpha _j\cdot \alpha _k\left(  \sigma _j\sigma _k
-\sigma _j \wedge \sigma_k  \right) \right] \right\}  \nonumber \\[3ex]
&& \times \exp \left\{ \sum_{j=1}^n\alpha _j\cdot \left[ \sigma _jx+\left(
1-\sigma _j\right) x_0\right] \right\}  \nonumber
\end{eqnarray}
This solves formally the heat equation with potential $-igV\left( x\right) $.
Usually, convergence properties are easier to handle for the heat equation
than for the Schr\"odinger equation. However, in our case it is the other
way around.

\begin{theorem}
Let $d=1$ and let $m$ be a positive measure on the Borel sets of ${\R}$ 
fulfilling condition (\ref{condition}). If $m\left( {\R \backslash}\left\{
0\right\} \right) >0$ then the power series in $g$%
\begin{eqnarray}
&&\ \sum_{n=0}^\infty \frac 1{n!}\left( \frac{-ig\left( t-t_0\right) }\hbar
\right) ^n\int_{\left[ 0,1\right] ^n}\int_{{\Ri}^n}\prod_{j=1}^n\mathrm{d}%
m\!\left( \alpha _j\right) \mathrm{d}^n\sigma  \\[3ex]
&& \times \exp \left\{ -\frac \hbar {2m}\left( t-t_0\right) \left[
\sum_{j=1}^n\sum_{k=1}^n\alpha _j\cdot \alpha _k\left(  \sigma _j\sigma _k
-\sigma _j \wedge \sigma_k   \right) \right] \right\}   \nonumber
\\[3ex]
&& \times \exp \left\{ \sum_{j=1}^n\alpha _j\cdot \left[ \sigma _jx+\left(
1-\sigma _j\right) x_0\right] \right\}   \nonumber
\end{eqnarray}
diverges for every $g\neq 0$ for any fixed $x_0,x,t_0<t$.
\end{theorem}

\TeXButton{Proof}{\proof}Either there exists $a_0>0$ with $m\left( \left[
a_0,\infty \right) \right) >0$ or $a_0<0$ with $m\left( \left( -\infty
,a_0\right] \right) >0$. Without lost of generality we assume $a_0>0$. We
use the shorthand 
\begin{eqnarray}
\lefteqn{F\left( \alpha _1,\ldots ,\alpha _n,\sigma _1,\ldots ,\sigma _n\right)} \\[3ex]
\ &:=&\exp \left\{ -\frac \hbar {2m}\left( t-t_0\right) \left[
\sum_{j=1}^n\sum_{k=1}^n\alpha _j\cdot \alpha _k\left(  \sigma _j\sigma _k
-\sigma _j \wedge \sigma_k   \right) \right] \right\}  \nonumber \\[3ex]
&& \times \exp \left\{ \sum_{j=1}^n\alpha _j\cdot \left[ \sigma _jx+\left(
1-\sigma _j\right) x_0\right] \right\}.  \nonumber
\end{eqnarray}
For $ \frac{3}{16} \leq \sigma_j , \sigma_k \leq \frac{4}{16} $ we have
\begin{equation}
  \sigma _j\sigma _k -\sigma _j \wedge \sigma_k   \leq -\frac{1}{16}. 
\end{equation}
Since the integrand is positive we get 
\begin{eqnarray}
\lefteqn{\left| \int_{\left[ 0,1\right] ^n}\int_{{\Ri}^n}F\left( \alpha _1,\ldots
,\alpha _n,\sigma _1,\ldots ,\sigma _n\right) \prod_{j=1}^n\mathrm{d}%
m\!\left( \alpha _j\right) \mathrm{d}^n\sigma \right|} \\[3ex]
\ &\geq &\int_{\left[ \frac 3{16},\frac 4{16}\right] ^n}\int_{\left[
a_0,\infty \right) ^n}F\left( \alpha _1,\ldots ,\alpha _n,\sigma _1,\ldots
,\sigma _n\right) \prod_{j=1}^n\mathrm{d}m\!\left( \alpha _j\right) \mathrm{d%
}^n\sigma  \nonumber \\[3ex]
\ &\geq &\int_{\left[ \frac 3{16},\frac 4{16}\right] ^n}\int_{\left[
a_0,\infty \right) ^n}\exp \left\{ \frac \hbar {2m}\left( t-t_0\right) \frac
1{16}n^2a_0^2\right\}  \nonumber \\[3ex]
&& \times \exp \left\{ -\sum_{j=1}^n\left| \alpha _j\right| \left( \left|
x_0\right| +\left| x\right| \right) \right\} \prod_{j=1}^n\mathrm{d}%
m\!\left( \alpha _j\right) \mathrm{d}^n\sigma  \nonumber \\[3ex]
\ &\geq &\left( \frac 1{16}\right) ^n\exp \left\{ \frac \hbar {32m}\left(
t-t_0\right) a_0^2n^2\right\} \left( \int_{a_0}^\infty \exp \left\{ -\left|
\alpha \right| \left( \left| x_0\right| +\left| x\right| \right) \right\} 
\mathrm{d}m\left( \alpha \right) \right) ^n.  \nonumber
\end{eqnarray}
By the above assumption for $a_0$ the last factor does not vanish. 
Thus we get for the series 
\begin{eqnarray*}
&& \!\!\!\!\!\!\!\!\!\! \sum_{n=0}^\infty \left| \frac 1{n!}\left( \frac{-ig\left( t-t_0\right) }%
\hbar \right) ^n\int_{\left[ 0,1\right] ^n}\int_{{\Ri}^n}F\left( \alpha
_1,\ldots ,\alpha _n,\sigma _1,\ldots ,\sigma _n\right) \prod_{j=1}^n\mathrm{%
d}m\!\left( \alpha _j\right) \mathrm{d}^n\sigma \right| \\[3ex]
\ &\geq& \sum_{n=0}^\infty \frac 1{n!}\left( \frac{\left| g\right| \left(
t-t_0\right) }{16\hbar }\int_{a_0}^\infty \exp \left\{ -\left| \alpha
\right| \left( \left| x_0\right| +\left| x\right| \right) \right\} \mathrm{d}%
m\!\left( \alpha \right) \right) ^n \\[3ex]
\ && \times  \exp \left( \frac \hbar {32m}\left( t-t_0\right)
a_0^2 n^2 \right)  \\[3ex]
\ &=&\infty .
\end{eqnarray*}

\hfill\TeXButton{End Proof}{\endproof}

\subsection{Time-dependent potentials\label{SuSeTimeDep}}

One of the advantages of the Feynman integral is that it can be easily
extended to time-dependent potentials.

\begin{theorem}
\label{ThPropTime}Let $m$ denote a complex measure on the Borel sets of ${\R}%
^d~\times ~\left[ t_0^{\prime },t^{\prime }\right] $; $d\geq 1$, such that 
\begin{equation}
\int_{{\Ri}^d}\int_{t_0^{\prime }}^{t^{\prime }}e^{C\left| \alpha \right| }\,%
\mathrm{d\!}\left| m\right| \!\left( \alpha ,\tau \right) <\infty ,\quad
\quad \forall \,\,C>0
\end{equation}
Then for $t_0^{\prime }\leq t_0<t\leq t^{\prime }$ 
\begin{equation}
I:=\sum_{n=0}^\infty \frac{\left( -i\right) ^n}{n!}\int_{\left[ t_0,t\right]
^n}\int_{{\Ri}^{dn}}I_0\cdot \prod_{j=1}^ne^{\alpha _j\cdot x\left( \tau
_j\right) }\prod_{j=1}^n\mathrm{d}m\!\left( \alpha _j,\tau _j\right) 
\end{equation}
exists as a generalized white noise functional in $(\mathcal{S}_d)^{-1}$.
The $T$-transform fulfills the following equation 
\begin{equation}
TI\left( \theta \right) =\sum_{n=0}^\infty \frac{\left( -i\right) ^n}{n!}%
\int_{\left[ t_0,t\right] ^n}\int_{{\Ri}^{dn}}T\left( I_0\cdot
\prod_{j=1}^ne^{\alpha _j\cdot x\left( \tau _j\right) }\right) \!\!\left(
\theta \right) \prod_{j=1}^n\mathrm{d}m\!\left( \alpha _j,\tau _j\right) 
\end{equation}
for all $\theta $ in a neighborhood of zero.
\end{theorem}

\begin{remark}
\emph{If we consider the Schr\"odinger equation for the whole class defined
in the above theorem, the potential becomes a distribution in the time
variable. This causes technical difficulties; we only use the
special forms given below.}
\end{remark}

\begin{theorem}
\label{ThPropSG}Let $m$ be as in Theorem \ref{ThPropTime}. If additionally $m
$ has either the special form 
\begin{equation}
\mathrm{d}m\!\left( \alpha ,\tau \right) =\sum_{j=1}^k\mathrm{d}m_j\left(
\alpha \right) \rho _j\left( \tau \right) \mathrm{d}\tau 
\end{equation}
with $k\in {\N}$, $m_j$ complex measures on the Borel sets of ${\R}^d$ and $%
\rho _j\in C^0\left( {\R,\C}\right) $ for all $j=1,\ldots ,k$;

\noindent or the special form 
\begin{equation}
\mathrm{d}m\!\left( \alpha ,\tau \right) =\rho \left( \alpha ,\tau \right) 
\mathrm{d}^d\alpha \,\,\mathrm{d}\tau 
\end{equation}
where $\rho :{\R}^d\times [t_0^{\prime },t^{\prime }]{\rightarrow \C}$ with 
$\rho \left( \alpha ,\cdot \right) $ continuous on $[t_o^{\prime
},t^{\prime }]$ for all $\alpha \in {\R}^d $%
 and $\sup_{\tau \in \left[ t_0^{\prime },t^{\prime }\right] }\left| \rho \left(
\alpha ,\tau \right) \right| $ in $L^1\left( {\R,}\mathrm{d}^d\alpha \right)$ 

\noindent
then the propagator 
\begin{eqnarray}
\lefteqn{K^{\left( \theta \right) }\left( x,t\mid x_0,t_0\right)}  \\[3ex]
 \  &=&\sum_{n=0}^\infty \frac{\left( -ig\right) ^n}{n!}\left( 2\pi i\left(
t-t_0\right) \right) ^{-d/2}\exp \left( ix_0\cdot \theta \left( t_0\right)
-ix\cdot \theta \left( t\right) \right)   \nonumber \\[3ex]
&& \times \int_{\left[ t_0,t\right] ^n}\int_{{\Ri}^{dn}}\exp \left\{
-\frac i2\int_{{t}_0}^t\left( \theta \left( s\right) +i\sum_{j=1}^n\alpha _j{%
\1}_{\left( \tau _j,t\right] }\left( s\right) \right) ^2\mathrm{d}s\right\} 
\nonumber \\[3ex]
&& \times \exp \left\{ -\frac 1{2i\left( t-t_0\right) }\left[
\int_{t_0}^t\theta \left( s\right) \mathrm{d}s+i\sum_{j=1}^n\alpha _j\left(
t-\tau _j\right) +\left( x-x_0\right) \right] ^2\right\}   \nonumber \\[3ex]
&& \times \exp \left( \sum_{j=1}^n\alpha _j\cdot x\right)
\prod_{j=1}^n\mathrm{d}\!m\left( \alpha _j,\tau _j\right)   \nonumber
\end{eqnarray}
solves the Schr\"odinger equation with potential $x \cdot \stackrel{_{\bullet }}{%
\theta }\left( t\right) +V\left( x,t\right) $ for all $t_0^{\prime
}<t_0<t<t^{\prime }$, where 
\begin{equation}
V\left( x,t\right) :=\sum_{j=1}^k\int_{{\Ri}^d}e^{\alpha \cdot x}\mathrm{d}%
m_j\left( \alpha \right) \rho _j\left( t\right) 
\end{equation}
or respectively 
\begin{equation}
V\left( x,t\right) :=\int_{{\Ri}^d}e^{\alpha \cdot x}\rho \left( \alpha
,t\right) \mathrm{d}^d\alpha .
\end{equation}
\end{theorem}

\noindent \textbf{Proof of Theorem \ref{ThPropTime} and \ref{ThPropSG}.
\thinspace \thinspace }The proof can be done in a similar way as in the
previous subsection. Only the measure $\mathrm{d}m\left( \alpha _j\right) 
\mathrm{d}\tau _j$ has to be replaced by $\mathrm{d}m\left( \alpha _j,\tau
_j\right) $. More explicitly we then get instead of bound (\ref
{EstimateIntegral}) the following 
\begin{eqnarray}
\lefteqn{\int_{\left[ t_0,t\right] ^n}\int_{{\Ri}^{dn}}C_n(\alpha _1,\ldots
,\alpha _j,\theta )\prod_{j=1}^n\mathrm{d\!}\left| m\right| \!\left( \alpha
_j,\tau _j\right)} \\[3ex]
\ &\leq &\left( 2\pi \left( t-t_0\right) \right) ^{-d/2}\exp \left( \left|
\theta \right| _0^2+\frac{\left| x-x_0\right| }{\sqrt{t-t_0}}\left| \theta
\right| _0\right)  \nonumber \\[3ex]
&& \times \,\left\{ \int_{{\Ri}^d}\int_{t_0}^t\exp \left( \left[ \left|
x-x_0\right| +\left| x_0\right| +2\sqrt{t-t_0}\left| \theta \right|
_0\right] \left| \alpha \right| \right) \,\,\mathrm{d\!}\left| m\right|
\!\left( \alpha ,\tau \right) \right\} ^n.  \nonumber
\end{eqnarray}
To complete the proof of Theorem \ref{ThPropSG} we have to justify the 
interchange of the time derivative with the $\alpha $%
-integrals. For this we use an extra condition as e.g. in the second special form the
integrability of $\sup_{\tau \in \left[ t_0^{\prime },t^{\prime }\right]
}\left| \rho \left( \alpha ,\tau \right) \right| $. \hfill%
\TeXButton{End Proof}{\endproof}}

\QSubDoc{Include morse}{
\LaTeXparent{C:/TKUNA/TEX;master}
\ChildStyles{amssymb} 
\ChildDefaults{chapter:0,page:1}

\section{A SPECIAL CASE: THE MORSE POTENTIAL \label{SeMorse}}

In order to illustrate the remarkable fact that the propagator is analytic
in the coupling constant we discuss the Morse potential as a
special case of the class of potentials we studied in Section \ref{SeUnbounded}.
This potential has been very useful in molecular and nuclear physics, see
for example. \cite{He50}

\begin{definition}
In $L^2\left( {\R},{\C}\right) $ we consider the Hamilton
operator 
\begin{equation}
H:=-\frac 12\triangle +g\left( e^{-2ax}-2\gamma e^{-ax}\right) 
\end{equation}
with $\gamma ,g,a\in $I\negthinspace R. As domain we choose 
\begin{equation}
D\left( H\right) :=C_0^\infty \left( {\R,}{\C}\right) 
\end{equation}
the set of infinite differentiable functions with compact support. For $g>0$
and $\gamma >0$ this is called the Morse potential.
\end{definition}

We now collect some well known results from operator theory. 
\cite{DuSc63,ReSi75}

\begin{proposition}
$H$ is symmetric and 
\[
D\left( H^{*}\right) =\left\{ f\in L^2\left( {\R,}{\C}%
\right) \cap C^1\left( {\R,}{\C}\right) \left| 
\begin{array}{c}
f^{\prime }\text{ is absolutely continuous } \\[3ex] 
\text{and }Hf\in L^2\left( {\R},{\C}\right) 
\end{array}
\right. \right\} 
\]
Since we have a real potential the deficiency indices are equal and thus
there exists a self-adjoint extension, which is not necessarily unique.
\end{proposition}

>From Theorem X.8, X.9 in Ref.~30 \nocite{ReSi75} we can derive the following proposition

\begin{proposition}
$H$ is essentially self-adjoint for $g\geq 0$ and it is not essentially
self-adjoint for $g<0$.
\end{proposition}

For the Morse potential ($g>0$) there exist treatments by algebraic, 
\cite{ChGuHa92} and by operator methods. \cite{NiSi79}\ Furthermore 
path integral techniques have been applied. 
\cite{Kl90,PaSo84,CaInWi83}. One uses path-dependent
space-time transformation to convert the path integral for the Morse
potential into the radial path integral for the harmonic oscillator in three
dimensions which is well known. Further refinements were done in Refs.~16 
\nocite{FiLeMu92} and 31. \nocite{Pe96} By latter technique one calculates the Green
function, the kernel of the resolvent, moreover one derives the spectrum and 
the eigenfunctions. Originally the method of path-dependent space-time 
transformation was applied to calculate the Feynman integral of the hydrogen 
atom. \cite{DuKl79,BlSi81}. The following formulas 
are taken from Ref.~16 \nocite{FiLeMu92} First we will have a look at the Green 
function. 
\begin{eqnarray}
\lefteqn{G\left( x^{\prime },x;E\right) :=\left\langle x^{\prime }\left| \left(
H-E\right) ^{-1}\right| x\right\rangle }  \label{EqGreen} \\[3ex]
&=&\frac{\Gamma \!\left( \left( 1+ \nu -
\gamma \omega \right)/2 \right) }{\omega {\left| a \right| }/2  \,\,
\Gamma \!\left(
 \nu+1\right) } exp{\left(\frac a2\left( x+x^{\prime
}\right)\right) }  \nonumber \\[3ex]
&&\!\!\!\!\!\!\!\!\!\times \left\{ \Theta \!\left( a(x-x^{\prime })\right)
 \,W_{ \gamma \omega/2, \nu/2 }\left( 
\omega \,e^{-ax^{\prime }}\right) M_{ \gamma
\omega/2 , \nu/2 } \left( 
\omega \,e^{-ax}\right) \right.   \nonumber \\[3ex]
&&\!\!\!\!\!\!\! \left. +\Theta \!\left( a(x^{\prime }-x)\right) \,M_{ \gamma 
\omega/2 ,\nu/2 }\left( \omega \, e^{-ax^{\prime }}\right) W_{ \gamma
\omega/2 , \nu/2 } \left( \omega \,e^{-ax}\right) \right\}   \nonumber
\end{eqnarray}
with $x^{\prime },x$ $\in $I\negthinspace R. $\Theta $ denotes the Heaviside
function with the convention $\Theta \left( 0\right) =1/2$, $\Gamma $
is Euler's gamma function, $M_{\varkappa ,\mu / 2}$ and $W_{\varkappa
,  \mu / 2}$ Whittaker's functions, \cite{Er53} $\omega :=2 \sqrt{2g}/
{\left| a \right |}$ and $\nu := 2 \sqrt{-2E}/{\left| a \right|}$.
The $\sqrt{\,\cdot \,}$
denotes the principal branch of the square root with the cut along the
negative real half line. Then we get for the spectrum 
\begin{eqnarray}
\sigma \left( H\right)  &=&\left[ 0,\infty \right)  \\[3ex]
&&\!\!\!\!\!\!\!\!\!\! \bigcup \left\{ -\frac 18a^2\left( \frac{2\gamma }{\left| a\right| }%
\sqrt{2g}-2n-1\right) ^2\left| n\in \text{I\negthinspace N}_0\text{ with }%
\frac{2\gamma }{\left| a\right| }\sqrt{2g}-2n-1>0\right. \right\}   \nonumber
\end{eqnarray}
and for the eigenvectors of the discrete eigenvalues 
\begin{eqnarray}
\Psi _n\left( x\right)  &=&\sqrt{\frac{\left| a\right| \left( \gamma 
\omega -2n-1\right) \Gamma \left( n+1\right) }{\Gamma
\left( \gamma \omega -n\right) }} \,\, 
\omega^{ \gamma \omega/2- \left(2n+1 \right)/2} \\[3ex]
&&\ \exp \left({-\frac{a}{2}\left( \gamma \omega  -\left(2n-1\right)\right)
x} \right) \exp \left ({-\frac {\omega}{2} e^{-ax}} \right) L_n^{\left( \gamma \omega
-2n-1\right) }\left( \omega e^{-ax}\right)   \nonumber
\end{eqnarray}
where $L_n^{\left( \mu \right) }$ is a generalized Laguerre polynomial. 
\cite{Er53}

\begin{proposition}
The Green function (\ref{EqGreen}), the eigenvectors and the discrete
eigenvalues are not analytic in $g$.
\end{proposition}

\TeXButton{Proof}{\proof}For the discrete eigenvalues the above statement is
obvious. The Green function is not even analytic in $\omega = 2 \sqrt{2g}
/{\left|a\right| } $. Using Ref.~35 \nocite{Bu69} we can rewrite (\ref{EqGreen}) as 
\begin{eqnarray*}
\  &&\!\!\!\!\!\! G\left( x^{\prime },x;E\right) =\frac{2\pi }{\left| a\right| \sin
\left( \nu \pi \right) }\exp \left( {-\frac \omega 2\left( e^{-ax^{\prime
}}+e^{-ax}\right) }\right) \\[3ex]
&&\ \times\left\{ \Theta \!\left( a(x-x^{\prime })\right) \frac{_1F_1\left( 
\left( 1+\nu -\gamma \omega \right)/2 ;1+\nu ,\omega e^{-ax}\right) }{\Gamma
\!\left( 1+\nu \right) }\right.  \\[3ex]
&&\ \quad \times \left[ -\omega ^\nu e^{-\nu a \left( x+x^{\prime }\right)/2 }%
\frac{\Gamma \!\left( \left( 1+\nu -\gamma \omega \right)/2 \right) }{%
\Gamma \!\left( \left( 1-\nu -\gamma \omega \right)/2 \right) }\frac{%
_1F_1\left( \left( 1+\nu -\gamma \omega \right)/2 ;1+\nu ,\omega
e^{-ax^{\prime }}\right) }{\Gamma \!\left( 1+\nu \right) }\right.  \\[3ex]
&&\ \quad \quad \left. +e^{\nu a \left( x^{\prime }-x\right)/2 }\frac{_1F_1\left(
\left( 1-\nu -\gamma \omega \right)/2 ;1-\nu ,\omega e^{-ax^{\prime
}}\right) }{\Gamma \!\left( 1-\nu \right) }\right]  \\[3ex]
&&\ +\Theta \!\left( a(x^{\prime }-x)\right) \frac{_1F_1\left( 
\left( 1+\nu -\gamma \omega \right)/2 ;1+\nu ,\omega e^{-ax^{\prime
}}\right) }{\Gamma \!\left( 1+\nu \right) } \\[3ex]
&&\ \quad \times \left[ -\omega ^\nu e^{-\nu a \left( x+x^{\prime }\right)/2 }%
\frac{\Gamma \!\left( \left( 1+\nu -\gamma \omega \right)/2 \right) }{%
\Gamma \!\left( \left( 1-\nu -\gamma \omega \right)/2 \right) }\frac{%
_1F_1\left( \left( 1+\nu -\gamma \omega \right)/2 ;1+\nu ,\omega
e^{-ax}\right) }{\Gamma \!\left( 1+\nu \right) }\right.  \\[3ex]
&&\ \quad \quad \left. \left. +e^{\nu a \left( x-x^{\prime }\right)/2 }\frac{%
_1F_1\left( \left( 1-\nu -\gamma \omega \right)/2 ;1-\nu ,\omega
e^{-ax}\right) }{\Gamma \!\left( 1-\nu \right) }\right] \right\} 
\end{eqnarray*}
with $\nu =2 \sqrt{-2E}/ {\left| a\right| }$ and $_1F_1$ denotes a
generalized hypergeometric function. \cite{Er53} The function 
\begin{equation}
\frac{_1F_1\left( a;b,x\right) }{\Gamma \left( b\right) }
\end{equation}
is entire in $a,b,x$. Thus we only have to investigate $\Gamma (
(1+\nu -\gamma \omega )/2)\omega ^\nu $. This is obviously not analytic in 
$ \omega $ near $ \omega =0 $. For
the eigenvectors we can proceed along the same line.\hfill%
\TeXButton{End Proof}{\endproof}

Although the Green function, the discrete eigenvalues and the eigenfunctions
are not analytic, the propagator has a perturbation series
which is uniformly absolutely convergent in the coupling constant for every
compact set in the variables $x,t,x_0,t_0$. For the Morse potential we
obtain an expansion for the propagator from (\ref{EqPropProduct}).
Putting $\theta =0$ and doing the $\mathrm{d}m\left( \alpha _j\right) $
integrations we get 
\begin{eqnarray}
\lefteqn{K\left( x,t\mid x_0,t_0\right)}  \\[3ex]
&=&K_0\left( x,t\mid x_0,t_0\right) \cdot \sum_{n=0}^\infty \frac{\left(
-ig\right) ^n}{n!}\left( t-t_0\right) ^n  \nonumber \\[3ex]
&&\times \sum_{j_1,\ldots ,j_n=1}^2\left( -2\gamma \right)
^{2n-\sum_{k=1}^nj_k}\int_{\left[ 0,1\right] ^n}\exp \left\{
-a\sum_{l=1}^nj_l\left( \sigma _lx+\left( 1-\sigma _l\right) x_0\right)
\right\}   \nonumber \\[3ex]
&&\times \exp \left\{ -\frac i2\left( t-t_0\right)
a^2\sum_{l=1}^n\sum_{k=1}^nj_kj_l\left[  \sigma _j\sigma _k
-\sigma _j \wedge \sigma_k   \right] \right\} \,\mathrm{d}^n\sigma.   \nonumber
\end{eqnarray}
In the sum over $n$ the coefficient of $g^n$ is bounded by 
\[
\frac 1{n!}\left| t-t_0\right| ^n\left( 1+2\left| \gamma \right| \right)
^ne^{2n\left| a\right| \left( \left| x-x_0\right| +\left| x_0\right| \right)
}.
\]

\noindent \textbf{Acknowledgments}

We are grateful to Professors Ch. Bernido, V. Bernido and Yu.G. Kondratiev
for fruitful discussion. We thank also our colleagues M. Grothaus and J.L.
da Silva for helpful comments. This work was supported in part by 
Financiamento Plurianual, JNICT, no. 219/94.

}
\newpage

\bibliographystyle{prsty}

\begin{thebibliography}{AlBrHa96}
\bibitem[AlBrHa96]{AlBrHa96}  Albeverio, S., Brze\TeXButton{z}{\'z}niak, Z.
and Haba, Z. (1996), \textit{On the Schr\"odinger Equation with potentials
which are Laplace transforms of measures.} Inst. Math. Univ. Bochum, SFB 237
Preprint Nr. \textbf{296}.

\bibitem[AlHK76]{AlHK76}  Albeverio\textbf{, }S. and H\o egh-Krohn, R.
(1976), \textit{Mathematical Theory of Feynman Path Integrals}. LNM \textbf{%
523,} Springer Verlag, Berlin, Heidelberg and New York.

\bibitem[Ba85]{Ba85}  Barroso, J.A. (1985), \textit{Introduction to
Holomorphy. }Mathematical Studies \textbf{106, }North-Holland, Amsterdam.%
\textbf{\ }

\bibitem[BeKo88]{BeKo88}  Berezansky, Yu.M. and Kondratiev, Yu.G. (1988), 
\textit{Spectral Methods in Infinite-Dimensional Analysis}, (in Russian),
Naukova Dumka, Kiev. English translation 1995, Kluwer Academic Publishers,
Dordrecht.

\bibitem[BlSi81]{BlSi81}  Blanchard, Ph. and Sirugue, M. (1981), \textit{%
Treatment of some singular potentials by change of variables in Wiener
integrals. }J.~Math.~Phys. \textbf{22}, 1372-1376.

\bibitem[BSST93]{BSST93}  Blanchard, Ph., Sirugue-Collin, M., Streit, L. and
Testard, D. (Eds., 1993), \textit{Dynamics of complex and Irregular Systems. 
}World Scientific, Singapore.

\bibitem[Bu69]{Bu69}  Buchholz, H. (1969), \textit{The Confluent
Hypergeometric Function.} Springer Verlag, Berlin, Heidelberg and New York.

\bibitem[CaInWi83]{CaInWi83}  Cai, P.Y., Inomata, A. and Wilson, R. (1983), 
\textit{Path-Integral Treatment of the Morse Oscillator,} Phys. Lett. 
\textbf{96} A, 117-120.

\bibitem[Ca60]{Ca60}  Cameron, R.H. (1960), \textit{A Family of Integrals
Serving to Connect the Wiener and Feynman Integrals. }J. Math. Phys. \textbf{%
39}, 126-140.

\bibitem[ChGuHa92]{ChGuHa92}  Chetouani, L., Guechi, L. and Hammann, T.F.
(1992), \textit{Algebraic Treatment of the Morse Potential. }Helv. Phys.
Acta \textbf{65}, 1069-1075.

\bibitem[CDLSW95]{CDLSW95}  Cunha, M., Drumond, C., Leukert, P., Silva, J.L.
and Westerkamp, W. (1995), \textit{The Feynman integrand for the perturbed
harmonic oscillator as a Hida distribution. }Ann. Physik\textbf{\ 4},\textbf{%
\ }53-67\textbf{.}

\bibitem[Do80]{Do80}  Doss, H. (1980),\textit{\ Sur une r\'esolution
stochastique de l`\'equation de Schr\"odinger \`a coefficients analytiques. }%
Comm. Math. Phys. \textbf{73}, 247-264.

\bibitem[DuSc63]{DuSc63}  Dunford, N., Schwartz, J.T. (1963), \textit{Linear
Operators.} Vol. II, Interscience Publishers, New York and London.

\bibitem[DuKl79]{DuKl79}  Duru, I.H. and Kleinert, H., (1979), \textit{%
Solution of the Path Integral for the H-Atom.} Phys. Lett. \textbf{84} B,
185-188.

\bibitem[Er53]{Er53}  Erd\'elyi, A. (Ed., 1953), \textit{Higher
Transcendental Functions. }The Bateman Manuscript Project, Vol. I, II,
McGraw-Hill, New York.

\bibitem[FPS91]{FPS91}  Faria, M., Potthoff, J. and Streit, L. (1991), 
\textit{The Feynman Integrand as a Hida Distribution}, J. Math. Phys\textbf{%
. 32},\textbf{\ }2123-2127.

\bibitem[FeHi65]{FeHi65}  Feynman, R.P. and Hibbs, A.R. (1965), \textit{%
Quantum Mechanics and Path Integrals. }McGraw-Hill, New York and London.

\bibitem[FiLeMu92]{FiLeMu92}  Fischer, W., Leschke, H. and M\"uller, P.
(1992), \textit{Changing dimension and time: two well-founded and practical
techniques for path integration in quantum physics. }J. Phys. A \textbf{25},
3835-3853.

\bibitem[GrLi68]{GrLi68}  Grauert, H., Lieb, I. (1968), \textit{%
Differential- und Integralrechnung III. }Springer Verlag, Berlin, Heidelberg
and New York.

\bibitem[GKSS96]{GKSS96}  Grothaus, M., Khandekar, D.C., Silva, J.L. and
Streit, L., (1996) \textit{The Feynman Integral for time dependent
anharmonic oscillators. }Madeira preprint 18/96, accepted to be published in
J. Math. Phys.

\bibitem[He50]{He50}  Herzberg, G. (1950), \textit{Molecular Spectra and
Molecular Structure, I. Spectra of Diatomic Molecules.} Van Nostrand
Reinhold Company, New York.

\bibitem[HKPS93]{HKPS93}  Hida, T., Kuo, H.H., Potthoff, J. and Streit, L.
(1993),\textit{\ White Noise. An infinite dimensional calculus}. Kluwer,
Dordrecht.

\bibitem[HS83]{HS83}  Hida, T. and Streit, L. (1983), \textit{Generalized
Brownian Functionals and the Feynman Integral.} Stoch. Proc. Appl. \textbf{16%
}, 55-69.

\bibitem[KS92]{KS92}  Khandekar, D.C. and Streit, L. (1992), \textit{%
Constructing the Feynman Integrand.} Ann. Physik \textbf{1}, 46-55.

\bibitem[Kl90]{Kl90}  Kleinert H. (1990), \textit{Path Integrals in Quantum
Mechanics, Statistics and Polymer Physics. }World Scientific, Singapore.

\bibitem[Ko78]{Ko78}  Kondratiev, Yu.G. (1978), \textit{Generalized
functions in problems of infinite dimensional analysis. }Ph.D. thesis, Kiev
University.

\bibitem[KLPSW96]{KLPSW96}  Kondratiev, Yu. G., Leukert, P., Potthoff, J.,
Streit, L. and Westerkamp, W. (1996), \textit{Generalized Functionals in
Gaussian Spaces: The Characterization Theorem Revisited. }J. Funct. Anal. 
\textbf{141}, 301-318.

\bibitem[KLS96]{KLS96}  Kondratiev, Yu.G., Leukert, P. and Streit,L. (1996), 
\textit{Wick Calculus in Gaussian Analysis}. Acta Appl. Math. \textbf{44},
269-294.

\bibitem[KoS93]{KoS93}  Kondratiev, Yu.G. and Streit, L. (1993), \textit{%
Spaces of White Noise distributions: Constructions, Descriptions,
Applications. I}. Rep. Math. Phys. \textbf{33}, 341-366.

\bibitem[Kuo92]{Kuo92}  Kuo, H.H. (1992), \textit{Lectures on White Noise
Analysis}. Soochow J. Math. \textbf{18}, 229-300.

\bibitem[Kuo96]{Kuo96}  Kuo, H.H. (1996), \textit{White Noise Distribution
Theory. }CRC Press, Boca Raton, New York, London and Tokyo..

\bibitem[LLSW93]{LLSW93}  Lascheck, A., Leukert, P., Streit, L. and
Westerkamp, W. (1993), \textit{Quantum Mechanical Propagators in Terms of
Hida Distributions}. Rep. Math. Phys. \textbf{33, }221-232.

\bibitem[Lu70]{Lu70}  Lukacs, E. (1970), \textit{Characteristic Functions.}
2nd edition,Griffin, London.

\bibitem[NiSi79]{NiSi79}  Nieto, M.M. and Simmons, L.M.Jr. (1979), \textit{%
Coherent states for general potentials. III. Nonconfining one-dimensional
examples.} Phys. Rev. D \textbf{20}, 1342-1350.

\bibitem[Ob94]{Ob94}  Obata, N. (1994), \textit{White Noise Calculus and
Fock Space}. LNM \textbf{1577}, Springer Verlag, Berlin, Heidelberg and New
York.

\bibitem[PaSo84]{PaSo84}  Pak, N.K. and Sokmen, I. (1984), \textit{General
new-time formalism in the path integral. }Phys. Rev. A \textbf{30},
1629-1635.

\bibitem[Pe96]{Pe96}  Pelster, A. (1996), \textit{Zur Theorie und Anwendung
nichtintegrabler Raum-Zeit-Transformationen in der klassischen Mechanik und
in der Quantenmechanik. }Ph.D. thesis, Universit\"at Stuttgart, Shaker
Verlag, Aachen.

\bibitem[Po91]{Po91}  Potthoff, J. (1991), \textit{Introduction to White
Noise Analysis}. Baton Rouge Preprint.

\bibitem[PS91]{PS91}  Potthoff, J. and Streit, L. (1991), \textit{A
characterization of Hida distributions.} J. Funct. Anal. \textbf{101},
212-229.

\bibitem[ReSi75]{ReSi75}  Reed, M. and Simon, B. (1975), \textit{Methods of
modern mathematical physics. }Vol. I, II, Academic Press, New York and
London.

\bibitem[Sch71]{Sch71}  Schaefer, H.H. (1971) \textit{Topological Vector
Spaces}. Springer Verlag,\textbf{\ }Berlin, Heidelberg and New York.

\bibitem[S93]{S93}  Streit, L. (1993), \textit{The Feynman Integral - Recent
Results}. In: \cite{BSST93}, 166-173\textbf{.}

\bibitem[SW93]{SW93}  Streit, L. and Westerkamp,W. (1993), \textit{A
Generalization of the Characterization Theorem for Generalized Functionals
of White Noise}. In: \cite{BSST93}, 174-187.

\bibitem[W93]{W93}  Westerkamp, W. (1993), \textit{A Primer in White Noise
Analysis}. In: \cite{BSST93}, 188-202.

\bibitem[W95]{W95}  Westerkamp, W. (1995), \textit{Recent Results in
Infinite Dimensional Analysis and Applications to Feynman Integrals. }Ph.D.
thesis, University of Bielefeld.

\end{thebibliography}
\LaTeXparent{C:/KUNA;master}
                      
\ChildDefaults{chapter:0,page:1}

\end{document}